\documentclass[prb,aps,epsf,twocolumn,showpacs,10pt,superscriptaddress
]{revtex4-1}
\usepackage{graphicx,amsfonts,times,bm,amsmath,verbatim,color,array}

\usepackage{natbib}
\usepackage{graphicx,amssymb,amsmath,ifthen,braket,xcolor}
\usepackage{bm}

\begin{document}


\title{Kitaev chain with a quantum dot}



\author{Chuanchang Zeng}
\affiliation{Department of Physics and Astronomy, Clemson University, Clemson, SC 29634, USA}

\author{Christopher Moore}
\affiliation{Department of Physics and Astronomy, Clemson University, Clemson, SC 29634, USA}

\author{Apparao M. Rao}
\affiliation{Department of Physics and Astronomy, Clemson University, Clemson, SC 29634, USA}

\author{Tudor D. Stanescu}
\affiliation{Department of Physics and Astronomy, West Virginia University, Morgantown, WV 26506, USA}

\author{Sumanta Tewari}
\affiliation{Department of Physics and Astronomy, Clemson University, Clemson, SC 29634, USA}


\date{\today}

\begin{abstract}
We solve analytically the problem of a finite length Kitaev chain coupled to a quantum dot (QD), which extends the standard Kitaev chain problem making it more closely related to the quantum dot-semiconductor-superconductor (QD-SM-SC) nanowire heterostructure that is currently under intense investigation for possible occurrence of Majorana zero modes (MZMs). Our analytical solution reveals the emergence of a robust  Andreev bound state (ABSs) localized in the quantum dot region as the generic lowest energy solution in the topologically trivial phase. By contrast, in the bare Kitaev chain problem such a solution does not exist. The robustness of the ABS in the topologically trivial phase is due to a partial decoupling of the component Majorana bound states (MBSs) over the length of the dot potential. As a result,  the signatures of the ABS in measurements that couple locally to the quantum dot, e.g., tunneling measurements, are identical to the signatures of topologically-protected  MZMs, which arise only in the topological superconducting (TS) phase of the Kitaev chain.
\end{abstract}

\pacs{}

\maketitle

\section{Introduction}

Non-Abelian Majorana zero modes (MZMs), \cite{read2000paired,kitaev2001unpaired,nayak2008non} which were theoretically predicted \cite{sau2010generic,tewari2010theorem,alicea2010majorana,sau2010non,lutchyn2010majorana,oreg2010helical,stanescu2011majorana} to arise as zero-energy excitations at the edges of low-dimensional spin-orbit coupled semiconductors with proximity induced superconductivity in the presence of a Zeeman field, have emerged \cite{stanescu2013majorana,beenakker2013search,elliott2015colloquium} as the leading candidate in the creation of topological quantum bits essential to fault-tolerant quantum computation. \cite{kitaev2001unpaired,nayak2008non} This research has been bolstered by recent experimental progress leading to observations of key signatures of Majorana zero modes \cite{mourik2012signatures,deng2012anomalous,das2012zero,rokhinson2012fractional,churchill2013superconductor,finck2013anomalous,albrecht2016exponential,deng2016majorana,zhang2017ballistic,chen2017experimental,nichele2017scaling,zhang2018quantized} in semiconductor-superconductor (SM-SC) nanowire heterostructures, particularly the emergence of a zero bias conductance peak in the tunneling conductance spectra at a finite magnetic field. Theoretically, such zero bias conductance peaks (ZBCPs)  were shown to also arise due to low energy states generated by several different phenomena unrelated to topology. \cite{bagrets2012class,liu2012zero,degottardi2013majoranaprl,degottardi2013majoranaprb,rainis2013towards,
adagideli2014effects,kells2012near,chevallier2012mutation,roy2013topologically,san2013multiple,ojanen2013topological,
stanescu2014nonlocality,cayao2015sns,klinovaja2015fermionic,san2016majorana,fleckenstein2017decaying,pikulin2012zero,
prada2012transport,liu2017andreev} However, the low-energy states of non-topological origins are usually found to generate ZBCPs that are not quantized at peak height $2e^2/h$ and/or are not stable against variations of various experimental control parameters such as magnetic field, chemical potential, and tunnel barrier height. This is the main reason why recent experiments capable of measuring ZBCPs which remain quantized at $2e^2/h$ over a finite range of control parameters, \cite{zhang2018quantized} as required theoretically for the signatures of topological MZMs, \cite{sengupta2001midgap,akhmerov2009electrically,law2009majorana,flensberg2010tunneling} have garnered a great deal of excitement. \cite{franz2018quantized}

To properly analyze the Majorana zero mode experiments on SM-SC heterostructures, it is useful to note that many of the systems under experimental investigation should be described as a quantum dot-semiconductor-superconductor (QD-SM-SC) nanowire heterostructure (rather than a simple SM-SC heterostructure without the QD as was originally proposed \cite{sau2010generic,tewari2010theorem,alicea2010majorana,sau2010non,lutchyn2010majorana,oreg2010helical,stanescu2011majorana}), because a QD is almost inevitably formed in the bare SM wire segment between the normal tunnel lead and the epitaxial SC shell owing to  band bending and/or disorder. \cite{deng2016majorana,zhang2018quantized} Therefore, while the topological properties of the theoretically proposed simple SM-SC heterostructure in the presence of spin-orbit coupling and Zeeman field can be described in terms of an effective model consisting of a finite length Kitaev chain, \cite{kitaev2001unpaired} the correct effective model for the systems under experimental investigation is a finite length Kitaev chain coupled to a QD, where the QD is defined by a region at the end of the chain in the presence of a local electric potential and vanishing superconducting pair potential $\Delta$. In this paper, we analytically solve this effective model in the long-wavelength, low-energy limit. In addition to providing the analytical solution to a valuable extension of the celebrated Kitaev model (i.e., Kitaev chain coupled to a quantum dot, an extension motivated by experiment), our study allows a qualitative understanding of recent numerical work \cite{moore2018two-terminal,moore2018quantized} on proximitized SM-SC heterostructures coupled to a QD, which has shown that it is possible to have quantized ZBCPs of height $2e^2/h$ forming robust plateaus with respect to the experimental control parameters even in the topologically trivial phase.

In this paper we analyze a Kiaev chain of length $L$, characterized by a superconducting pair potential $\Delta$, which is coupled to an end QD of length $x_0$ (see Eq.~\ref{eq:hamiltonian2}) defined by an effective potential of height $V$ ($V=0$ in the bulk of the Kitaev chain). Experimentally, the effective potential  in the QD region (which we model, for simplicity, as a step-like potential of height $V$)  may be induced by a local gate and/or by a position-dependent work function difference between the SM and the SC (which is nonzero in the proximitized segment of the wire and vanishes in the uncovered regions).
Note that this type of position-dependent effective potential is manifestly different from the smooth confinement potential at the end of the chain considered in Ref.~[\onlinecite{kells2012near}] (see also Ref.~[\onlinecite{moore2018two-terminal}]). More importantly, the mechanisms for the formation of robust near-zero-energy non-topological ABSs are qualitatively different in the two models.
 In particular, in the presence of a smooth confinement potential\cite{kells2012near,moore2018two-terminal} the pair of component MBSs constituting a robust  near-zero-energy ABS originates from two \textit{different} spin channels of a confinement-induced sub-band, while in the presence of a step-like potential (in the QD region), with either positive $V$ (i.e. potential barrier) or negative $V$ (potential well),\cite{moore2018two-terminal,moore2018quantized} the component MBSs originate from the \textit{same} spin channel. This is why the topological properties of the QD-SM-SC hybrid structure with a step-like potential \cite{moore2018two-terminal,moore2018quantized} can be understood in terms of an effective representation of a Kitaev chain  coupled to a QD  (since the low-energy physics involves a single spin channel), while the SM-SC heterostructure with smooth confinement potential \cite{kells2012near}  cannot be analyzed using such a representation (because in this case both spin channels are required) .


First, we solve analytically the Hamiltonian for a finite length bare Kitaev chain (i.e. without the quantum dot) and obtain the wave functions corresponding to the lowest energy eigenvalues. To the best of our knowledge, the lowest energy wave functions with eigenvalues $\pm \epsilon$ emerging in the ``topological'' phase of a finite length Kitaev chain, with the putative Majorana energy eigenvalues $\pm \epsilon$ oscillating with the chemical potential and the chain length $L$, have so far only been estimated perturbatively based on the overlap of end-localized wave functions corresponding to a semi-infinite Kitaev chain or SM-SC Majorana wire.\cite{Sarma_2012} By contrast, our analytical treatment of the finite chain provides non-perturbative solutions for the energy splitting oscillations of the putative Majorana modes (as function of the chain length and chemical potential), as well as  the exponential decay of the amplitude of these oscillations with increasing system size. In particular, we show explicitly that the energy splitting oscillations are a direct consequence of imposing appropriate boundary conditions in a finite system.
Next, armed with these solutions, we solve the problem of a finite length Kitaev chain coupled to a quantum dot, where by quantum dot we mean a small region at the end of the chain defined by a local potential ``step'' of height $V$ and a reduced (possibly vanishing) induced superconducting pair potential, as suggested by the experimental setups involving semiconductor-superconductor hybrid structures.\cite{deng2016majorana,zhang2018quantized}
Our analytical solution of the full problem is characterized by a pair of robust low energy Bogoliubov-de Gennes (BdG) states with energies $\pm \epsilon$ localized in the quantum dot region as the generic lowest energy eigenstates in the topologically trivial phase of the Kitaev chain.
We emphasize that  no such near-zero-energy robust BdG states exist as low energy solutions in the topologically trivial phase of the finite length Kitaev chain \textit{without} a QD. In the topological superconducting phase, the lowest energy solutions are (topological) MZMs localized at the two ends of the chain.
We find that the robustness of the near-zero-energy BdG states $\phi_{\pm \epsilon}$ that emerge in the presence of the QD is due to a partial decoupling of the component Majorana bound states (MBSs), $\chi_A=\frac{1}{\sqrt{2}}\left[\phi_{\epsilon}+\phi_{-\epsilon}\right]$ and $\chi_B=\frac{i}{\sqrt{2}}\left[\phi_{\epsilon}-\phi_{-\epsilon}\right]$, over the length of the quantum dot. It follows that such partially-separated ABSs (ps-ABSs), which were first introduced in the numerical study of the SM-SC heterostructure coupled to a QD, \cite{moore2018two-terminal,moore2018quantized} generate signatures in experiments involving local probes, e.g., in charge tunneling experiments, identical to the signatures of topological MZMs.

The reminder of this article is organized as follows. In Section II we provide some preliminaries for the Kitaev chain model with periodic boundary conditions, which is applicable for infinitely long systems. In Section III, we detail the non-perturbative solution of the finite length Kitaev chain (with open boundary conditions). In Section IV, we  solve the problem of a finite length Kitaev chain coupled to a QD both analytically and numerically (for comparison). First, in Section IV A, we consider the case of no proximity effect in the QD from the adjoining SC, and find that, in this case, near-zero-energy ABSs \textit{do not} exist as low energy solutions in the topologically trivial phase of the Kitaev wire. In Section IV B, we assume a slice of the QD adjoining the SC to be proximitized and show that correct matching of the boundary conditions in the different regions of the Kitaev chain coupled to the QD produces robust near-zero-energy ABSs localized in the QD region as generic low energy solutions in the topologically trivial phase of the bulk Kitaev chain. We also analyze the wave functions of the component MBSs of the low energy ABSs and find that these states are spatially separated by the length of the proximitized region in the QD. We discuss the overlap of the component MBSs and the resultant splitting oscillations of the so-called partially separated ABSs, and find that the splitting is generically lower for these states because of the existence of the adjoining Kitaev chain in which the component MBSs can relax. We end with a summary of the main results and some concluding remarks in Section V.

\section{Kitaev Model Preliminaries}
The one-dimensional model of topological superconductivity proposed by Kitaev \cite{kitaev2001unpaired} can be derived from the tight binding Hamiltonian for a 1D superconducting wire as follows,
\begin{equation}
H=-\sum\limits_{j=1}^{N}\left[\mu c_j^\dagger c_j -\left(tc_{j+1}^\dagger c_j-\Delta c_j c_{j+1}+ h.c.\right)\right]\label{eq:TBHamm}
\end{equation}
where $t$, $\mu$, and $\Delta$ are the nearest neighbor hopping amplitude, chemical potential, and superconducting pairing potential, respectively, and $c$ and $c^{\dagger}$ are the second quantized creation and annihilation operators. Introducing the operators $\gamma_{2j-1}=c_{j}+c^{\dagger}_{j}$ and $\gamma_{2j}=i(c^{\dagger}_{j}-c_{j})$ allows the Hamiltonian $H$ to be rewritten as,
\begin{align}
\begin{split}
H&=-i\frac{\mu}{2}\sum\limits_{j=1}^{N}\gamma_{2j-1}\gamma_{2j}\\
&+\frac{i}{2}\sum\limits_{j=1}^{N-1}\left[(t+|\Delta|)\gamma_{2j}\gamma_{2j+1}+(-t+|\Delta|)\gamma_{2j-1}\gamma_{2j+2}\right]\label{eq:kitChain}
\end{split}
\end{align}
In the limit $\mu=0$ and $t=|\Delta|>0$ the Hamiltonian becomes,
\begin{equation}\label{eq:H2}
H=it\sum\limits_{n=1}^{N-1}\gamma_{2n}\gamma_{2n+1}.
\end{equation}
Because $\gamma_1$ and $\gamma_{2N}$ do not appear in the Hamiltonian this represents the topological phase of the wire described in Eq.~\ref{eq:kitChain}, in which a single pair of zero energy MZMs appear at the ends of the wire while the bulk of the wire remains gapped at an energy of $\pm\left|t\right|$. More generally, applying periodic boundary conditions, and Fourier transforming the Hamiltonian in Eq.~\ref{eq:TBHamm} into momentum space, the Bogoliubov-de Gennes (BdG) Hamiltonian can be written as (with the lattice constant $a=1$),
\begin{align}
\begin{split}
&H=\int dx\Psi^{\dagger}\left(k\right)H_{BdG}\Psi\left(k\right)\hspace{10mm}\Psi^\dagger =\left(c^\dagger,c\right)
\\
&H_{BdG}=(-2t\cos k-\mu)\tau_z+2\Delta\tau_y \sin k
\end{split}
\label{eq:HamKit}
\end{align}
where $k$ is the momentum and $\tau_z,\tau_y$ are the Pauli matrices operating in the particle-hole space. The bulk band structure for the wire, found by diagonalizing Eq.~\ref{eq:HamKit} is $E=\sqrt{(2t\cos k+\mu)^2+4\left|\Delta\right|^2\sin^2 k}$, which shows a bulk band gap closure at $k=\{0,\pi\}$ for $\mu=\pm2t$, representing the topological quantum phase transition (TQPT) as described in the Kitaev model.

In the long wavelength limit near the band gap closure as $k\to 0$, such that $\sin k \to k$ and $\cos k \to \left(1-k^2/2\right)$, Eq.~\ref{eq:HamKit} can be rewritten as,
\begin{equation}
\tilde{H}_{BdG}=(-t\partial^2_x-\tilde{\mu})\tau_z+i \tilde{\Delta} \partial_x\tau_y\label{eq:Hamiltonian}
\end{equation}
where $\tilde{\mu}=\mu-2t$ and $\tilde{\Delta}=2\Delta$.
Because the chemical potential is being measured from the bottom of the band ($\tilde{\mu}=\mu-2t$), the phase transition points in Eq.~\ref{eq:Hamiltonian} ($\tilde{\mu}=\pm 2t$) are now at $\mu=0$ and $\mu=4t$.

Solutions to the eigenvalue equation
\begin{equation}
\tilde{H}_{BdG}\phi\left(x\right)=E\phi\left(x\right)
\label{eq:eigenvalueEquation}
\end{equation}
are found by applying a trial wave function of the form $\phi \left(x\right)=\left(u\left(x\right) ,v\left(x\right)\right)^T=\tilde{\phi}\left(x\right) \left(\tilde{u} ,\tilde{v}\right)^T$ in which the spatial dependence is fully incorporated in the function $\tilde{\phi} \left(x\right)\propto e^{ i \lambda x}$ with $\tilde{u},\tilde{v}$ being independent of $x$. Substituting the trial function $\phi(x)$ in Eq.~\ref{eq:eigenvalueEquation} we find the characteristic equation,
\begin{equation}
E^2-\left(\lambda ^2-\mu\right)^2 -\Delta^2 \lambda^2 =0
\label{eq:charEquation}
\end{equation}
along with the following constraint on the spinor degrees of freedom
\begin{equation}
\tilde{v}= i \frac{\lambda^2-\mu-E}{\Delta \lambda} \tilde{u}\label{eq:constraint}.
\end{equation}
Here, all the terms are written in terms of the hopping energy $t$, so all the parameters with dimension of energy are rendered dimensionless in the rest of the paper.

Note that $\lambda$ in the trial wave function $\tilde{\phi}(x)$ is a complex number which can be written as, $\lambda=k+i q$ ($k,q \in \mathcal{R}$). Substituting it back to Eq.~\ref{eq:charEquation} gives us,
\begin{equation}
\begin{split}
\left (E^2-\Delta^2 (k^2-q^2)+4 q^2 k^2-(k^2-q^2-\mu)^2 \right) \\
 -2 i q k\left( 2(k^2-q^2)+\Delta^2-2\mu \right)=0
\end{split}
\label{eq:charEquation_1}
\end{equation}
The eigen-energy $E$ being real, it follows that the imaginary part in Eq.~\ref{eq:charEquation_1} must vanish,
\begin{equation}
\begin{split}
-2 i q k\left( 2(k^2-q^2)+\Delta^2-2\mu \right)=0
\end{split}
\label{eq:charEquation_2}
\end{equation}
There are three possible cases which can be extracted from Eq.~\ref{eq:charEquation_2}, namely, (a) $q = 0$, and $\lambda=k$ is purely real, (b) $k = 0$ and $\lambda=iq$ is purely imaginary, and (c) $k,q \neq 0$ with  $\lambda=k+iq$ a complex number. Substituting the three solutions to Eq.~\ref{eq:charEquation_2}, namely, $q=0$, $k=0$, and $2(k^2-q^2)=(2\mu-\Delta^2)$ back in Eq.~\ref{eq:charEquation_1} for the cases (a), (b), and (c), respectively, we have,
\begin{subequations}
    \begin{align}
        &(k^2-\mu)^2+\Delta^2 k^2-E^2 \equiv 0\\
        &(q^2+\mu)^2-\Delta^2 q^2-E^2\equiv 0 \\
     &(\Delta^2 \mu-\Delta^4/4)-4k^2 q^2-E^2\equiv 0
     \end{align}
\label{eq:charEquation_3}
\end{subequations}
 We can now solve $k,q$ in a form $k(E),q(E)$ (energy dependent) for each case in Eq.~\ref{eq:charEquation_3}. However, we can roughly analyze the approximate range of the eigen-energy $E$ in each case before moving on. In case (a), $E^2=(k^2-\mu)^2+\Delta^2 k^2 $ would give us $|E| \geq |\mu|$ if $\mu \leq 0$. As for $\mu > 0$ we have $E^2 \approxeq \mu^2+(\Delta^2-2\mu)k^2$ for $k^2 \ll \mu$, which would again give us $|E|$ with some value near $\mu$. Similarly in case (b), we have $E^2 \approxeq \mu^2 +(2\mu -\Delta^2) q^2$ for $q^2  \ll \mu$. It follows that both cases (a) and (b) cannot support a low energy solution ($E \ll |\mu|$) appropriate for MZMs in the topological phase. However, in case (c), we can have a solution with low energy-eigenvalue $E$, which is our main interest. In case (c) we have $E^2=\Delta^2(\mu-\Delta^2/4)-4k^2q^2$, which could be tuned to get a near-zero-energy solution independently of $k,q$ in the topological phase. It will indeed provide us with a nontrivial solution  in terms of putative Majorana zero modes as discussed below.

\section{Finite length Kitaev chain and splitting oscillations}\label{sec:finiteLenSplitting}
\begin{figure}
	\begin{center} \includegraphics[width=0.48\textwidth]{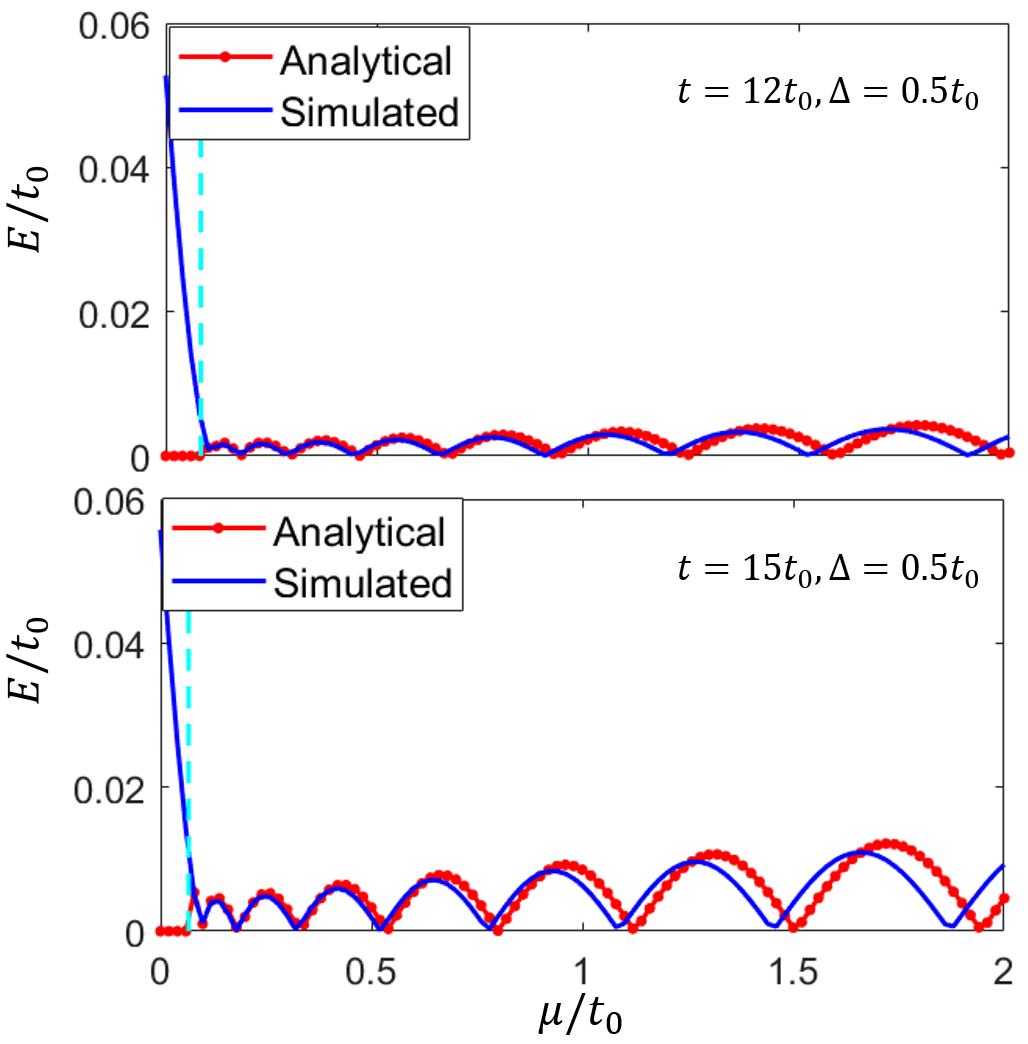}\llap{\parbox[b]{162mm}{\large\textbf{(a)}\\\rule{0ex}{55mm}}}\llap{\parbox[b]{162mm}{\large\textbf{(b)}\\\rule{0ex}{12mm}}}
		\includegraphics[width=0.48\textwidth]{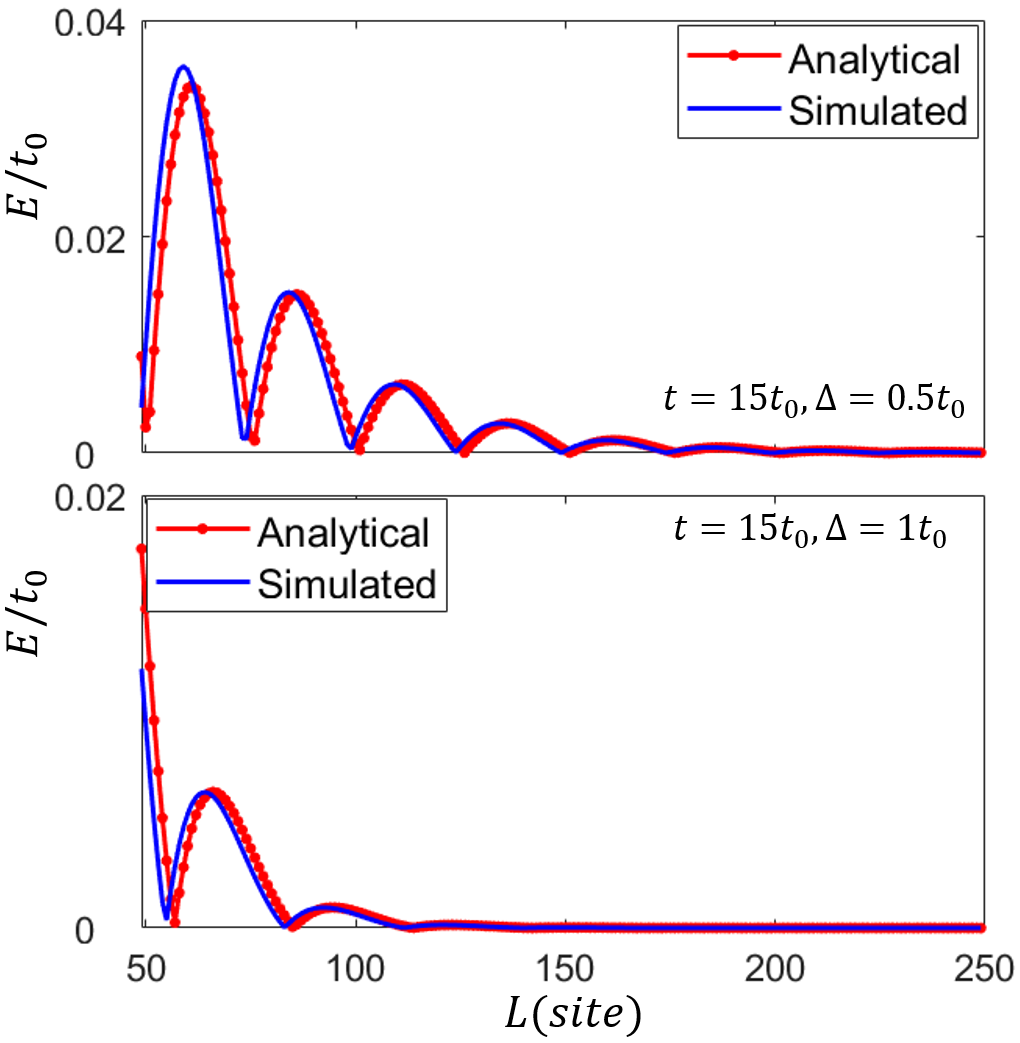}\llap{\parbox[b]{162mm}{\large\textbf{(c)}\\\rule{0ex}{55mm}}}\llap{\parbox[b]{162mm}{\large\textbf{(d)}\\\rule{0ex}{12mm}}}
	\end{center}
	\caption{(Color online) Lowest-energy spectra for the finite length Kitaev chain in the topological phase $\mu > 0$ as a function of chemical potential $\mu$ in (a)--(b) and Length $L$ in (c)--(d). Analytical results (red dotted line) are based on Eq.~\ref{eq:energyDecay}, while the simulation results (blue solid line) are from direct diagonalization of the tight binding hamiltonian in Eq.~\ref{eq:TBHamm} for a finite length $L$. The cyan dashed lines in (a)--(b) show the condition $\mu \geq\Delta ^2/4$ discussed below Eq.~\ref{eq:energyDecay}, after which the analytical solutions are valid. It shows a lower hopping energy $t$ reduces the energy splitting in (a) relative to that in (b); a higer SC $\Delta$ offers a more effective exponential protection in (d) than that in (c). All of this can be explained by the dependence of the lowest energy $E$ on the parameters $\mu, L$, respectively, through Eq.~\ref{eq:energyDecay}. A slight shift of the analytical results relative to the simulation results is caused by the dropped terms for $q$ and $k$ in Eq.~\ref{eq:energyDecay}. The other model parameters are $L=71$ in (a)--(b) and $\mu=0.25 t_0$ in (c)--(d), and $t_0=1$.}	
	\label{fig:energyKitaev}	
\end{figure}
In order to find the solutions to the full problem of a finite length Kitaev chain coupled to a QD, we first analytically solve the finite length bare Kitaev chain \textit{without} the QD. To the best of our knowledge, these solutions, which reveal the exponential decay and splitting oscillations of the lowest energy eigenvalues with the chain length $L$ and the chemical potential $\mu$ non-perturbatively, have not been written before. Later, we will use these exact solutions to find the solutions for the full problem of the Kitaev chain coupled to a quantum dot by matching the wave functions of the full Bogoliubov-de Gennes equations.

For a given one-dimensional Kitaev chain in the topological superconducting phase of finite length $L$, to support a pair of low energy solutions at energies $\pm E$, the roots $\lambda_i$ of Eq.~\ref{eq:charEquation} are necessarily complex (case (c) below Eq.~\ref{eq:charEquation_2}). combining Eq.~\ref{eq:charEquation_3}(c) with $2(k^2-q^2) = (2\mu-\Delta^2)$ (requirement for a complex $\lambda$, discussed in the last section), solutions of the form $\lambda\equiv iq + k $ are found in which
\begin{align}
\begin{split}
&k=\frac{1}{\sqrt{2}}\left(\left(\mu-\Delta^2/2\right)+\sqrt{\mu^2-E^2}\right)^{\frac{1}{2}}
\\
&q=\frac{1}{\sqrt{2}}\left(-\left(\mu-\Delta^2/2\right)+\sqrt{\mu^2-E^2}\right)^{\frac{1}{2}}
\end{split}
\label{eq:waveVector}
\end{align}
When combined with the constraint on the spinor degrees of freedom in Eq.~\ref{eq:constraint}, and the assumption that $\left|E\right|<\left|\mu\right|$, the general eigenfunction solution for the Hamiltonian in Eq.~\ref{eq:Hamiltonian} can be constructed as the linear combination $\phi(x)=\sum_{i=1} ^4 c_i \phi_i \left(x\right)$ such that
\begin{align}
\begin{split}
&\phi_1(x)= e^{i \lambda x}
\left( \begin{array}{cc} 1 \\ -q \alpha +i k \beta \end{array} \right) \tilde{u},
\\
&\phi_2(x)= e^{-i \lambda x}
\left( \begin{array}{cc} 1 \\ q \alpha -i k \beta \end{array} \right) \tilde{u},
\end{split}
\label{eq:eigen vector}
\end{align}
with $\phi_3\left(x\right)=\phi^\ast_1\left(x\right)$ and $\phi_4\left(x\right)=\phi^\ast_2\left(x\right)$. The two energy dependent weight components $\alpha, \beta$ are defined as,
\begin{equation}
\alpha\equiv\frac{1}{\Delta} \left(1+\sqrt{\frac{\mu+E}{\mu-E}}\right)\text{;}\hspace{2mm}\beta\equiv\frac{1}{\Delta} \left(1- \sqrt{\frac{\mu+E}{\mu-E}}\right)
\label{eq:alphaBeta}
\end{equation}
The energy $E$ can be found by constructing a matrix equation $\mathcal{A} X=0$ such that $\mathcal{A}$ consists of the four wave functions $\phi_i\left(x\right)$ with the boundary conditions $\phi_{\tilde{u}} \left(x=0\right)=\phi_{\tilde{v}} \left(x=0\right)=\phi_{\tilde{u}} \left(x=L\right)=\phi_{\tilde{v}} \left(x=L\right)=0$ applied, here $\phi_{\tilde{u}},\phi_{\tilde{v}} $ are the spinor components of $\phi(x)$ and $X=(c_1,c_2,c_3,c_4)^T$. The existence of nontrivial solution for $X$ requires $\mathcal{D} et(\mathcal{A})=0$, which yields the transcendental equation
\begin{equation}
k|\beta| \sinh{q L}= q \alpha|\sin{k L}|
\label{eq:determinant}
\end{equation}
Because we are interested in the lowest energy modes such that $E\ll\mu$, Eq.~\ref{eq:waveVector} is expanded to first order in $E/\mu$, $q \approxeq q_{F}+\mathcal{O}(E^2)$ and $k \approxeq k_{F} +\mathcal{O}(E^2)$, resulting in
\begin{equation}
\begin{split}
q_{F}=\Delta/2, \hspace{5mm} & k_F=(\mu -\left(\Delta/2\right)^2)^{\frac{1}{2}}
\end{split}
\label{eq:fermiVectors}
\end{equation}
Similar expansion of Eq.~\ref{eq:alphaBeta} yields
\begin{equation}
\frac{|\beta|}{|\alpha|} \approxeq \frac{E}{2\mu}\label{eq:weightAprox}
\end{equation}
Combining Eqs.~\ref{eq:fermiVectors}-\ref{eq:weightAprox} with Eq.~\ref{eq:determinant} and solving for $E$, we analytically find the exponentially protected ground state energy solution for a finite 1D $p$-wave superconducting nanowire
\begin{equation}
E\approxeq \frac{4 \mu q_F}{k_F} e^{-q_{F}L}| \sin (k_{F} L)| +\mathcal{O}(e^{-3q_{F} L})
\label{eq:energyDecay}
\end{equation}

Results following from Eq.~\ref{eq:energyDecay} are plotted in Fig.~\ref{fig:energyKitaev} (dotted lines) and compared with those of a direct numerical diagonalization of the Hamiltonian in Eq.~\ref{eq:Hamiltonian} (solid lines). Here we note that because $q$ and $k$ shown in Eq.~\ref{eq:waveVector} are real, the above solution is valid for energy values which are not near the TQPT point, such that $0\leq E^2\leq\left(\mu\Delta^2-\Delta^4/4\right)$, resulting in $\mu\geq\Delta^2/4$. The cyan line in Fig.~\ref{fig:energyKitaev}(a)-(b) shows this critical value of the chemical potental $\mu$, above which the analytical and simulated results are in close agreement.
\begin{figure}
	\begin{center}
		\includegraphics[width=0.46\textwidth]{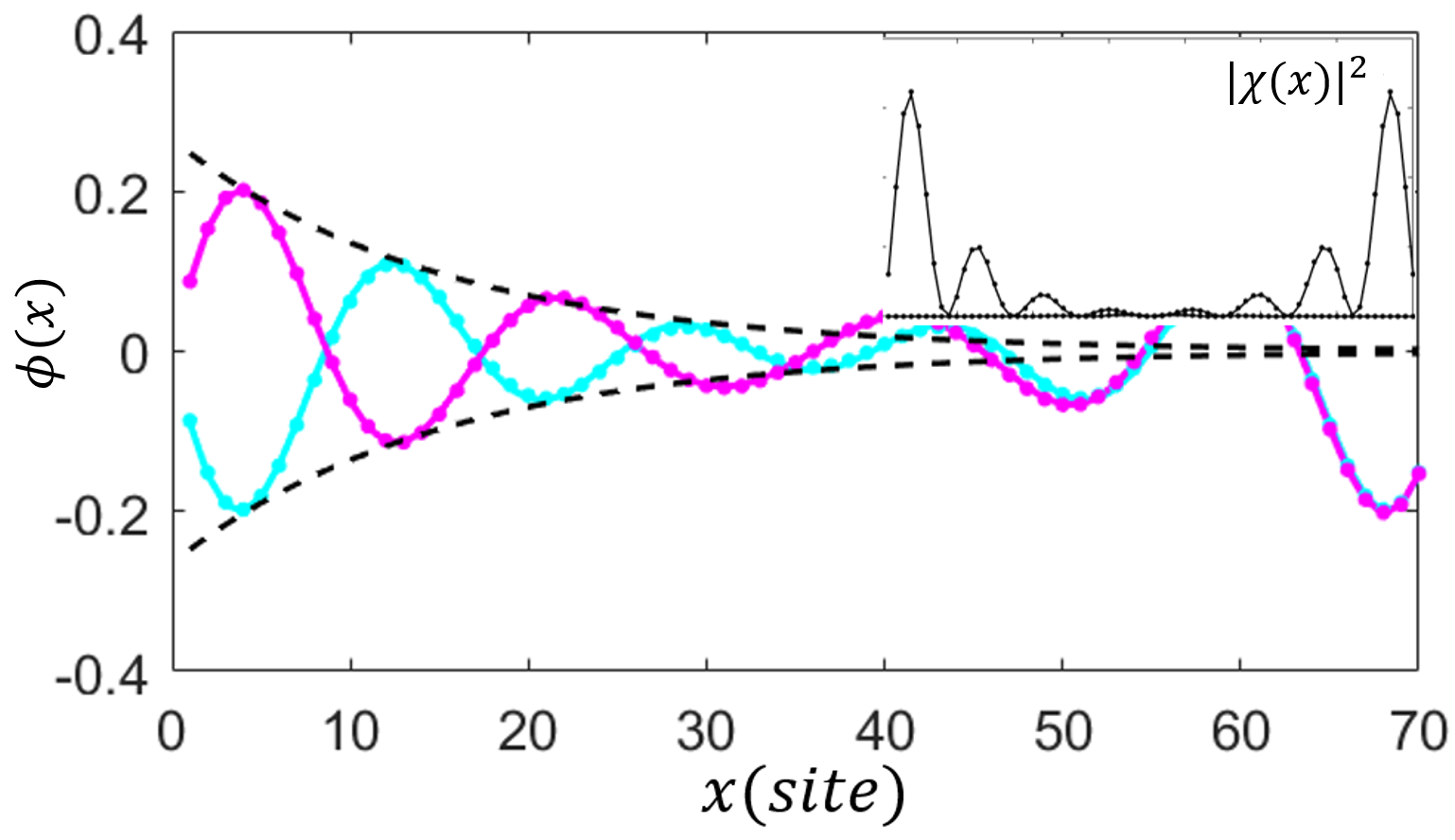}
	\end{center}
	\caption{(Color online) Lowest energy wave functions $\phi_{\pm E}$ as defined in Eq.~\ref{eq:waveFunction} and its counterpart by particle-hole transformation, for the lowest energy modes of a pure Kitaev chain in the topological regime ($\mu > 0$). The inset shows the corresponding MZMs as defined by Eq.~\ref{eq:waveModes}. The dotted lines correspond to analytical results while the solid lines correspond to numerical simulation. The black dashed line shows the exponential decay envelope of the wave function which is proportional to $e^{-q_F L }$. Parameters used were $\Delta=t_0$, $t=15t_0$, and $\mu=2t_0$. In the topological phase the putative MZM wave functions are localized at the two ends of the chain as shown in the inset. No such near-zero-energy subgap state exists as low energy solution in the non-topological phase of the Kitaev chain \textit{without} the quantum dot.}
	\label{fig:WFpure}
\end{figure}
Because the Hamiltonian as shown in Eq.~\ref{eq:eigenvalueEquation} is real, the non-degenerate eigenfunctions $\phi(x)$ associated with this Hamiltonian must be either purely real or purely imaginary, resulting in $c_{1,2}=\pm c_{3,4}$. In the limit $E\rightarrow 0$, the weight coefficients in Eq.~\ref{eq:alphaBeta} are $q\alpha\simeq 1$ and $\beta\simeq 0$, hence the spinor part for wavefunction $\phi(x)$ can be written as $\left(\tilde{u},\tilde{v}\right)^T=\left(1,-sign(q)\right)^T$(the spinor term $\tilde{u}$ is incorporated into the normalization factor $c_1, c_2$). After applying the boundary conditions $\phi\left(0\right)=\phi\left(L\right)=0$, solutions to the eigenvalue equation Eq.~\ref{eq:eigenvalueEquation} can be found of the form,
\begin{equation}
\begin{split}
\phi(x)&=c_1 e^{-q_F x}\sin{k_F x}\left(\begin{array}{cc} 1 \\ -sign\left(q\right)\end{array} \right) \\
&+c_2 e^{q_F\left(x-L\right)}\sin{\left(k_F\left(L-x\right)\right)} \left(\begin{array}{cc} 1 \\ sign\left(q\right)\end{array} \right)
\end{split}
\label{eq:waveFunction}
\end{equation}
where $c_1$ and $c_2$ are normalization coefficients and $\sin{\left(k_F\left(L-x\right)\right)}$ is taken to satisfy the boundary condition at $x=L$. Because $k_F$ and $q_F$ in Eq.~\ref{eq:waveFunction} are derived from $k$ and $q$  to first order in $E/\mu$ as seen in Eq.~\ref{eq:fermiVectors}, the two terms in $\phi\left(x\right)$ will not simultaneously equal to zero at the boundaries $x=0$ and $x=L$, but will when the full expressions for $k$ and $q$ are used.

Due to the particle-hole symmetry ($\tau_x^{\dagger}\tilde{H}_{BdG}\tau_x =-\tilde{H}_{BdG}$) of the Hamiltonian in Eq.~\ref{eq:eigenvalueEquation}, if $\phi_E\left(x\right)=\left(u\left(x\right), v\left(x\right)\right)^T$ is a solution to the BdG equation with energy $E$, then $\phi_{-E}\left(x\right)=\left(v^*\left(x\right), u^*\left(x\right)\right)^T$ is also a solution with energy $-E$. From these solutions linear combinations of the form, $\chi_A=\frac{1}{\sqrt{2}}\left(\phi_E\left(x\right)+\phi_{-E}\left(x\right)\right)$ and $\chi_B=\frac{i}{\sqrt{2}}\left(\phi_E\left(x\right)-\phi_{-E}\left(x\right)\right)$, are constructed representing a pair of partially overlapping MBSs. The BdG states $\phi_{\pm E}(x)$ described in Eq.~\ref{eq:waveFunction} are represented as a pair of partially overlapping MBSs of the form,
\begin{equation}
\begin{split}
&\chi_A= \tilde{c}_1 e^{-q_F x}\sin{k_F x}\left( \begin{array}{cc} i \\ -i \end{array} \right)
\\
&\chi_B= \tilde{c}_2 e^{q_F (x-L)}\sin{(k_F (L-x))} \left( \begin{array}{cc} 1 \\ 1 \end{array} \right)
\end{split}
\label{eq:waveModes}
\end{equation}
where $\tilde{c}_1,\tilde{c}_2$ are the normalization coefficients.  Though the Majorana wave functions $\chi_A,\chi_B$ defining bound states at the left and right ends are not exact eigenstates of the BdG Hamiltonian for the finite length Kitaev chain, they are useful in describing the interpolation of a low energy ABS into a pair of MBSs. Fig.~\ref{fig:WFpure} shows analytical results (dotted lines) based on Eq.~\ref{eq:waveFunction}, \ref{eq:waveModes} in close agreement with numerical results (solid lines). The left and right MBSs $|\chi_{A/B}|^2$ are spatially protected due to exponential decay (black dashed lines) of the wave functions. Note the boundaries from analytical results now are modified to be consistent with that from numerical simulation, where the boundary condition for the first and last site in TBM is not well defined. We find no  near-zero-energy subgap state as low energy solution in the non-topological phase of the Kitaev chain \textit{without} the quantum dot.

\section{Finite length Kitaev Chain attached to a Quantum Dot}\label{sec:finiteLenKit}
\begin{figure}
	\begin{center}
		\includegraphics[width=0.48\textwidth]{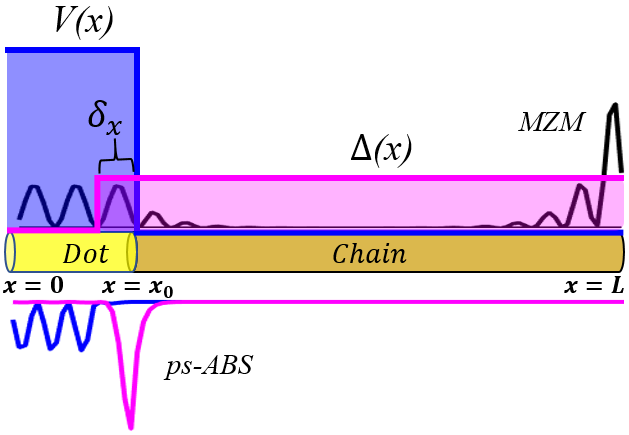}
	\end{center}
	\caption{(Color online)
		Schematic of a finite length Kitaev chain in which a fraction of the chain is not covered by the superconductor. This part of the chain (yellow) with vanishing superconducting pair potential $\Delta(x)$, and an effective electric potential $V(x)$ which may be induced by tunnel gates, is called a quantum dot. A proximitized region within the quantum dot with finite length $\delta_x$ is introduced. We also show the wave functions of the topological MZMs and the partially separated ABSs (in which the component MBSs are spatially separated over the length of the quantum dot), which are the generic lowest energy excitations in the topological and the non-topological phases of the Kitaev chain, respectively.}\label{fig:schematic}
\end{figure}
The one-dimensional finite length Kitaev chain with a quantum dot attached at the left end of the wire, schematically shown in Fig.~\ref{fig:schematic}, can be modeled with the Hamiltonian,
\begin{align}
\begin{split}
\tilde{H}_{BdG}&= -\left(\partial_x^2+\mu-V\left(x\right)\right)\tau_z-i\Delta\left(x\right)\partial_x\tau_y
\end{split}\label{eq:hamiltonian2}
\end{align}
for which $V\left(x\right) = V\Theta\left(x_0-x\right)$ in which $V$ can be positive (representing a potential barrier) or negative (representing a potential well) and $\Delta\left(x\right) = \Delta\Theta\left(x-x_0+\delta_x\right)$ in which $x_0$ is the length of the QD, and $\delta_x$ is the length of the proximitized region within the quantum dot (shown in Fig.~\ref{fig:schematic}), caused by the adjacent superconductor. Looking for $E\rightarrow0$ eigen-energy solutions we consider the eigenvalue equation given as Eq.~\ref{eq:eigenvalueEquation}. Uncoupling the near-zero-energy wave function solutions gives,
\begin{subequations}
	\begin{align}
    \left(-\partial ^2_x  + \Delta(x) \partial_x -\mu+V\left(x\right)  -E\right)f & =0\label{eq:secondOrderPDEsA} \\
    \left(-\partial ^2_x - \Delta(x) \partial_x -\mu+V\left(x\right)  -E\right)g & =0\label{eq:secondOrderPDEsB}
    \end{align}\label{eq:secondOrderPDEs}%
\end{subequations}
in which $f=u+v$ and $g=u-v$, where $u(x)$ and $v(x)$ are the spinor components of $\phi(x)$. The uncoupled equations in Eq.~\ref{eq:secondOrderPDEs} are equivalently valid for the coupled BdG equation given by Eq.~\ref{eq:eigenvalueEquation} in the near-zero-energy limit ($E \rightarrow 0$). In the limit that the proximitized region within the quantum dot goes to zero $\left(\delta_x\rightarrow 0\right)$ solutions to Eq.~\ref{eq:secondOrderPDEs} can be written as,
\begin{align}
\begin{split}
f(x)&=f_0(x) \Theta(x_0-x)+f_1(x)\Theta(x-x_0)\\
g(x)&=g_0(x) \Theta(x_0-x)+g_1(x)\Theta(x-x_0)
\end{split}\label{eq:dotTrivial}
\end{align}
where $f_0$ and $g_0$ represent the wave functions within the dot region and $f_1$ and $g_1$ are wave functions within the Kitaev-chain (the case $\delta_x \neq 0$ is discussed in Sec.~\ref{sec:finiteProx}). The Equations for $f(x)$ and $g(x)$ in Eq.~\ref{eq:secondOrderPDEs} are identical except for a change in sign of the superconducting term $\Delta\left(x\right)$. Thus if a solution to $g\left(x\right)$ is found, the corresponding wavefunction $f\left(x\right)$ can be inferred using the relation $f\left(x \right)\propto g\left(-x + \textit{d} x \right)$ in which $\textit{d} x$ is a constant shift. 

Below we first consider the case where the quantum dot has no proximitized region with non-zero superconducting pair potential adjacent to the SC interface, followed by the case where there is a slice of proximitized region within the quantum dot of width $\delta_x$. From our analytical solutions we find that, in the absence of a proximitized region within the QD, there are no robust low energy ABS solutions in the topologically trivial phase, whereas topological MZMs do appear in the topological superconducting phase of Kitaev chain. The low energy partially separated ABSs, on the other hand, are the generic lowest energy solutions localized in the quantum dot in the presence of a slice of proximitized region of width $\delta_x$ adjacent to the SC interface.
\subsection{No Proximity Coupling Within the QD}
\begin{figure}
	\begin{center}
		\includegraphics[width=0.46\textwidth]{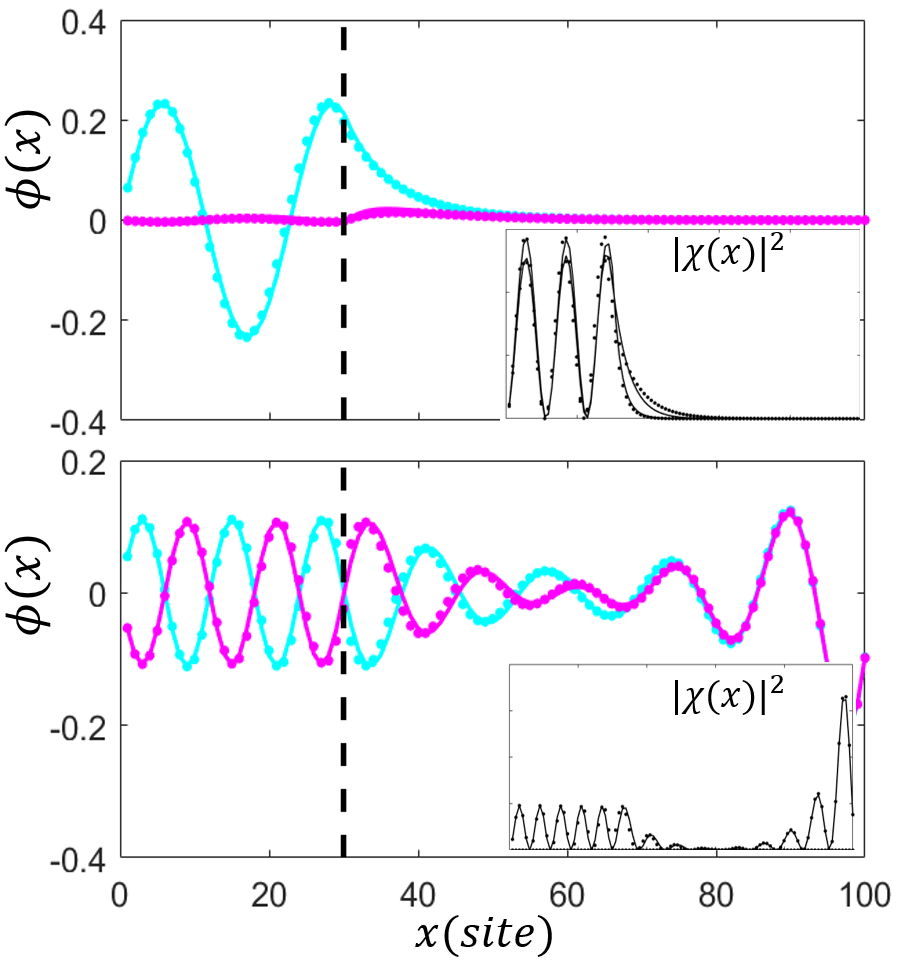}
		\llap{\parbox[b]{170mm}{\large\textbf{(a)}\\\rule{0ex}{79mm}}}
		\llap{\parbox[b]{170mm}{\large\textbf{(b)}\\\rule{0ex}{40mm}}}
	\end{center}
	\caption{(Color online) Wave functions for the lowest energy modes of a QD-Kitaev chain without a proximitized region in the QD ($\delta_x=0$ in Fig.~\ref{fig:schematic}) (a) corresponding to Eq.~\ref{eq:dotTrivialVars} for the topologically trivial regime with $\mu=-0.5t_0$, showing a pair of BdG wave functions $\phi_{\pm\varepsilon}$ in which the constituent MBSs are sitting directly on top of each other (see inset); and (b) corresponding to Eq.~\ref{eq:dotTopVars} for the topological regime with $\mu=3.5t_0$ in which a pair of MZMs are separated by the length of the wire (see inset). The insets show the MBSs associated with the low energy BdG wave functions. The dotted lines show the analytical results while the solid lines are from  numerical simulations. Because the constituent MBSs are strongly overlapping in (a), there is no robust near-zero-energy ABS in the topologically trivial phase in the absence of a proximitized region ($\delta_x = 0$) in the QD. The black dashed lines mark the QD-SC boundary at $x=x_0$. Parameters used were $V=-1.5t_0$, $\Delta=t_0$ and $t=10t_0$.}
	\label{fig:waveFunctions}
\end{figure}

We first consider the case for which the length of the proximitized region within the QD is zero $\left(\delta_x=0\right)$.  Assuming a topologically trivial state $\left(\mu <0\right)$ within the bulk of the Kitaev chain, and a potential well in the QD region $\left(V\left(x\right)<0\right)$ the effective chemical potential in the QD is $\left(\mu-V\left(x\right)\right) \gtrapprox 0$. Under these conditions the solutions to the Eq.~\ref{eq:secondOrderPDEs} for the entire QD-Kitaev chain can be written as
\begin{align}
\begin{split}
f_0(x)&=g_0(x)=c_0\sin(k_0x)\\
g_1(x)&=c_1 (e^{(-\lambda_0 +k_1)x} -e^{2k_1 L-(\lambda_0 +k_1)x})\\
f_1 (x)&=c^\prime_1\left(e^{(-\lambda_0 +k_1)\left(2L-x\right)}-e^{\left(k_1+\lambda_0\right)x-2\lambda_0 L}\right)
\end{split}\label{eq:dotTrivialVars}
\end{align}
with $k_0\equiv\sqrt{|\mu-V+E|}$, $k_1\equiv\sqrt{\left(\Delta/2\right)^2-\left(\mu+E\right)}$, and $\lambda_0 = q_F$ (defined from Eq.~\ref{eq:secondOrderPDEs}). Here the wave vector $k_1$ appearing in the definition of $f_1$ and $g_1$ in Eq.~\ref{eq:dotTrivialVars} describes the topologically trivial state within the Kitaev chain, and thus is not the same as $k$ previously defined for the topological state in Eq.~\ref{eq:waveVector}. For a potential barrier within the QD region $\left(\mu-V\left(x\right)<0\right)$ as opposed to a quantum well $\left(\mu-V\left(x\right)>0\right)$ the $\sin\left(k_0x\right)$ term as defined in Eq.~\ref{eq:dotTrivialVars} can be replaced by $\sinh\left(k_0x\right)$. The coefficients $c_0$, $c_1$, and $c_1^{\prime}$ are found by applying the boundary conditions $g_0(x)|_{x_0} = g_1(x)|_{x_0}$, $f_0(x)|_{x_0} = f_1(x)|_{x_0}$, $g^{\prime}_0(x)|_{x_0}=g^{\prime}_1(x)|_{x_0}$, and $f^{\prime}_0(x)|_{x_0}=f^{\prime}_1(x)|_{x_0}$ resulting in the energy dependent transcendental equations,
\begin{align}
\begin{split}
\lambda_0&=-k_0\cot{k_0 x} - k_1\coth{k_1\left(L-x_0\right)}\\
\lambda_0&=k_0\cot{k_0 x} + k_1\coth{k_1\left(L-x_0\right)}
\end{split}\label{eq:trans}
\end{align}
through which the lowest energy $E$ can be found numerically. Note that Eq.~\ref{eq:trans} produces two solutions for $E$, and we take the lower one as the lowest eigen-energy $E$. Once we know the eigen-energy $E$, the quantities $k_0,k_1$ can be derived from the expressions given below Eq.~\ref{eq:dotTrivialVars}, and so are the wave functions in Eq.~\ref{eq:dotTrivialVars}.
For the case in which the Kitaev chain is in the topological phase ($\mu >0$), wave functions of the form
\begin{align}
\begin{split}
f_0(x)&=g_0(x)=a_0 \sin(k_0x)\\
g_1(x)&=a e^{-q_F(x-x_0)}\sin(k_F(x-L))\\
f_1 (x)&=a ^{\prime} e^{-q_F(L-x)}\sin(k_F(L-x))
\end{split}\label{eq:dotTopVars}
\end{align}
can be found, where $a_0$, $a$, and $a^{\prime}$ are normalization factors. Note that wave functions as in Eq.~\ref{eq:waveFunction} are used for the topological chain here. The $\sin(k_0 x)$ term within the dot region can again be replaced with $\sinh\left(k_0 x\right)$ for values of the chemical potential such that $(\mu-V)<0$. The wavefunction within the Kitaev chain is expected to be of the same format as that of the pure Kitaev chain in topological phase. Matching the boundary conditions at $x=x_0$ for $g_{0},g_{1}$ and $f_{0},f_{1}$ respectively gives
\begin{equation}
\begin{split}
k_0\cot(k_0 x_0)&=-q_F+k_F\cot(k_F(x_0-L))\\
k_0\cot(k_0 x_0)&=q_F-k_F\cot(k_F(L-x_0))
\end{split}\label{eq:transtop}
\end{equation}
The above two equations are effectively equivalent for $q_F\ll k_F\cot\left(k_F\left(L-x_0\right)\right)$. As before, we can take the lower energy solution from Eq.~\ref{eq:transtop} as the eigen-energy $E$, and then the wave functions given in Eq.~\ref{eq:dotTopVars} can be derived. Analytical solutions for the topologically trivial (Eq.~\ref{eq:dotTrivialVars}) and topological (Eq.~\ref{eq:dotTopVars}) lowest energy wave functions of a finite QD-Kitaev chian are shown in Fig.~\ref{fig:waveFunctions} (dotted lines) to be in close agreement with numerical results (solid lines). The sinusoidal wave within the dot region and the exponentially decaying wave over the Kitaev chain are shown in Fig.~\ref{fig:waveFunctions}(b) with the black dashed line marking the boundary between the QD and the Kitaev chain. In Fig.~\ref{fig:waveFunctions}(a) inset, the constituent MBSs are sitting directly on top of each other, resulting in the absence of a robust near-zero-energy ABS in the topologically trivial phase of the Kitaev chain with no proximitized region in the QD ($\delta_x=0$ in Fig.~\ref{fig:schematic}).
\subsection{Finite Proximitized Region Within the QD}\label{sec:finiteProx}
\begin{figure}
	\begin{center}
		\includegraphics[width=0.46\textwidth]{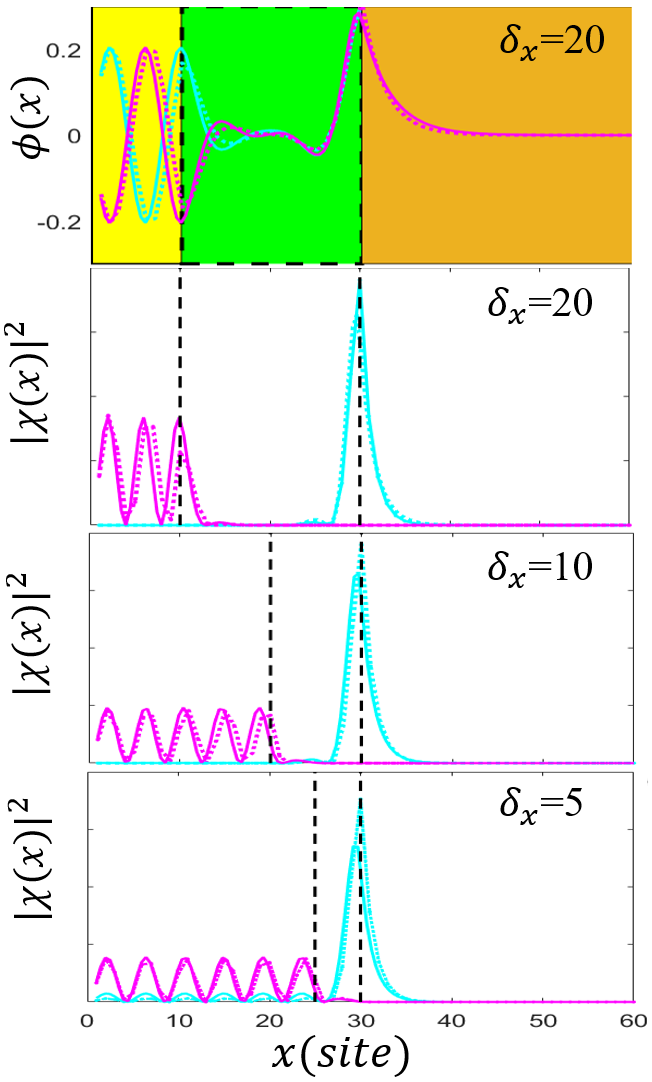}
		\llap{\parbox[b]{17mm}{\large\textbf{(a)}\\\rule{0ex}{122mm}}}
		\llap{\parbox[b]{17mm}{\large\textbf{(b)}\\\rule{0ex}{88mm}}}
		\llap{\parbox[b]{21mm}{\large\textbf{(c)}\\\rule{0ex}{50mm}}}
		\llap{\parbox[b]{21mm}{\large\textbf{(d)}\\\rule{0ex}{12mm}}}
	\end{center}
	\caption{(Color online) Wave functions for the lowest energy mode of a Kitaev chain coupled to a QD where the QD contains a proximitized region of finite length $\delta_x$ (see Fig.~\ref{fig:schematic})). (a) Wave functions for the lowest energy modes within the bare QD (yellow), the proximitized part of the QD (green), and the bulk Kitaev chian (orange), plotted using analytical results based on Eq.~\ref{eq:absWavefunction} (dotted lines) and direct numerical solutions using a tight binding Hamiltonian (solid lines). Here, the parameters are such that the proximitized region of the QD satisfies $\mu-V>0$, with $V=-2.5t_0$, while the Kitaev chain is topologically trivial with $\mu=-1.25t_0$. (b)-(d) Spatial profiles of the component pair of MBSs of a partially separated ABS for the proximitized regions of various lengths $\delta_x$. (b) shows the MBSs corresponding to the wave functions in (a). Samples are taken for values corresponding to the crossed diamonds in Fig.~\ref{fig:overlap}(c). The MBSs are separated on the order of the length of the proximitized region $\delta_x$ within the QD, marked by the black dotted lines. The figures illustrate that the ps-ABSs form essentially because the proximitized region in the QD satisfies the effective chemical potential $\tilde{\mu} > 0$, partially decoupling the ABS into a pair component MBSs, which are then spatially separated by the width of the proximitized region.
	Here the dot length is $x_0=30$, the total length of the QD-Kitaev chain is $L=100$, the hopping energy is $t=2.5 t_0$, and superconducting pairing potential is $\Delta=t_0$.}\label{fig:wfDot}
\end{figure}
Now we consider the case in which a finite proximitized region forms within the quantum dot adjacent to the SC interface $\left(\delta_x>0\right)$. In this case, the solutions to $\tilde{H}_{BdG}\phi\left(x\right)=E\phi\left(x\right)$, associated with Eq.~\ref{eq:hamiltonian2} are found by dividing the QD-Kitaev chain system into three regions as shown in Fig.~\ref{fig:schematic}): a pure quantum dot $\left(\Delta=0,V\neq 0\right)$, a finite proximitized region within the QD located near the QD-SC boundary $\left(\Delta\neq 0,V\neq 0\right)$, and a finite length Kitaev chain $\left(\Delta\neq 0,V =0\right)$. As before we assume that the chemical potential $\mu$ within the bulk of the Kitaev chain is $\mu\lessapprox 0$ such that the chain is in the topologically trivial phase. We also assume a potential well $\left(V(x) <0\right)$ within the QD region. It follows that the effective chemical potential within the proximitized region of the QD satisfies $\tilde{\mu}\equiv\left(\mu-V (x) \right)\gtrapprox 0$. Under these conditions we will use a sinusoidal wave function $g_0(x)$ in the region covered by the pure QD, the wave function given in Eq.~\ref{eq:waveFunction} for the proximitized region within the QD (call it $g_p(x)$, with ``$p$'' indicating solution valid in the proximitized region), and the wave function $g_1(x)$ appropriate for topologically trivial phase within the Kitaev chain,
\begin{equation}
g(x)=
\begin{cases}
a_0\sin\left(k_0 x\right),  & (g_0\left(x\right)) \\
a_{p} e^{-\lambda_0 x}\sin(k_{p} x+\delta \phi),   &(g_{p}(x))\\
a_1(e^{(-\lambda_0+k_1)x}-e^{2 k_1 L-(\lambda_0+k_1)x}),  &(g_1(x))
\end{cases}
\label{eq:absWavefunction}
\end{equation}
in which $a_0$, $a_{p}$, and $a_1$ are normalization factors, $k_0$, $k_1$, and $\lambda_0$ are as defined earlier, and $k_{p}=\sqrt{\left(\mu+E-V\right)-\left(\Delta/2\right)^2}$. A phase factor $\delta \phi$ is introduced for $g_{p}(x)$ because there are no fixed boundary values for the region $x\in [\left(x_0-\delta_x\right),x_0]$. Matching the boundary conditions at $x=x_0-\delta_x$ for $g_0 (x)$ and $g_{p} (x)$ and at $x=x_0$ for $g_{p} (x)$ and $g_1 (x)$ will  result in a pair of energy dependent transcendental equations given below which can be solved numerically for $E$ and $\delta\phi$.
\begin{equation}
\begin{split}
\lambda_0+k_0\cot(k_0( x_0-\delta_x))&= k_{p}\cot(k_{p}(x_0-\delta_x)+\delta \phi)\\
k_{p}\cot(k_{p}x_0+\delta \phi)&=-k_1 \coth{\left(k_1(L-x_0)\right)}
\end{split}\label{transcendental groups}
\end{equation}
As before, once the eigen-energy $E$ is known, wave vectors $k_0, k_p, k_1$ could be derived as well. The coefficients ($a_0, a_p, a_1 $) for the wave functions in Eq.~\ref{eq:absWavefunction} are then found by substituting the values $E, \delta \phi$ back into the boundary value equations. The term $e^{-\lambda_0 x}\sin(k_{p}x+\delta\phi)$ for the proximitized region will show a pair of spatially separated MBSs which are separated by the length of the proximitized region forming inside the QD.
\begin{figure}
	\begin{center} \includegraphics[width=0.48\textwidth]{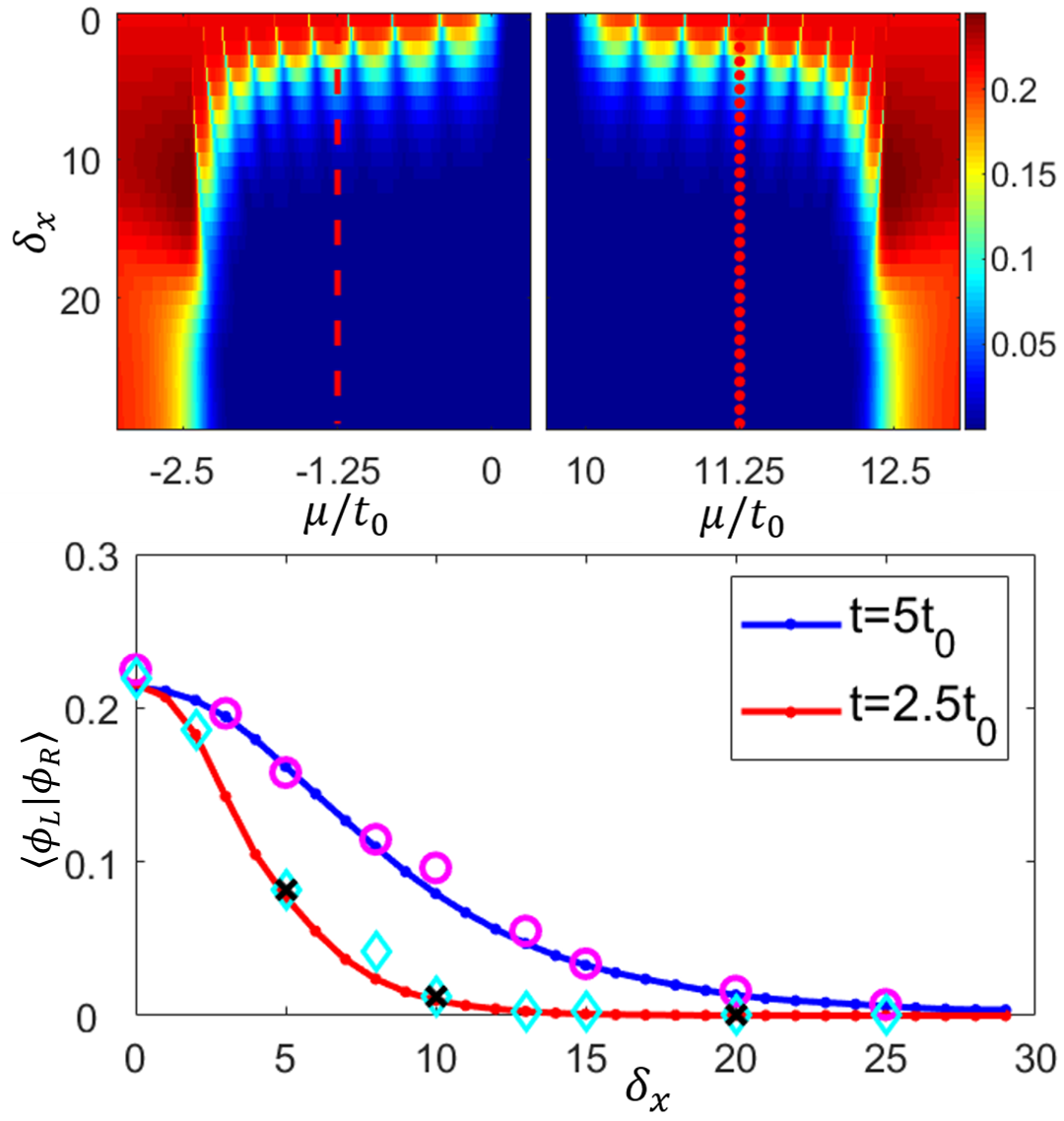}\llap{\parbox[b]{142mm}{\Large\textbf{\textcolor{white}{(a)}}\\\rule{0ex}{82mm}}}\llap{\parbox[b]{26mm}{\Large\textbf{\textcolor{white}{(b)}}\\\rule{0ex}{82mm}}}\llap{\parbox[b]{143mm}{\Large\textbf{(c)}\\\rule{0ex}{38mm}}}
	\end{center}
	\caption{(Color online) {Overlap between the left and right MBSs $\braket{\phi_L|\phi_R }$ (defined as $\braket{\phi_L|\phi_R }=\int dx |\phi_L||\phi_R|$) of a partially separated ABS, as a function of the width of the proximitized region  $\delta_x$ within the QD, and chemical potential $\mu$, for a potential well (a) and a potential barrier (b) in the QD region. Since the mechanism for the formation of the ps-ABS involves the effective chemical potential ($\tilde{\mu}=\mu-V$) in the proximitized part of the QD being in the topological regime, the potential well in the QD region ($V<0$) works for the bare $\mu<0$, while the potential barrier in the QD region works for the bare $\mu > 4t$.  As the length of the proximitized region  $\delta_x$ within the QD approaches zero, the overlap between the left and right MBSs dramatically increases (red). For finite values of the length of the proximitized region $\delta_x\geq 5$ minimal overlap between the left and right MBSs (blue) can be seen within the topologically trivial regime, $\mu\in\{-2.5t_0,0\}$ (a) and $\mu\in\{10t_0,12.5t_0\}$ (b)  supporting the presence of ps-ABSs. Here $\delta_x=30$, $L=100$, and $t=|V|=2.5t_0$. (c) Overlap between the left and right MBSs as a function of the length of the proximitized region within the dot. We find the same red curve corresponding to vertical line cuts taken from (a) and (b). The blue curve represents results for the same calculation as in the red curve but with a different hopping energy $t$. The cyan diamonds and the magenta circles represent the analytical results. The wave functions corresponding to the three black crossed diamonds are given in Fig.~\ref{fig:wfDot}.}
	}\label{fig:overlap}	
\end{figure}

\begin{figure}
	\begin{center} \includegraphics[width=0.48\textwidth]{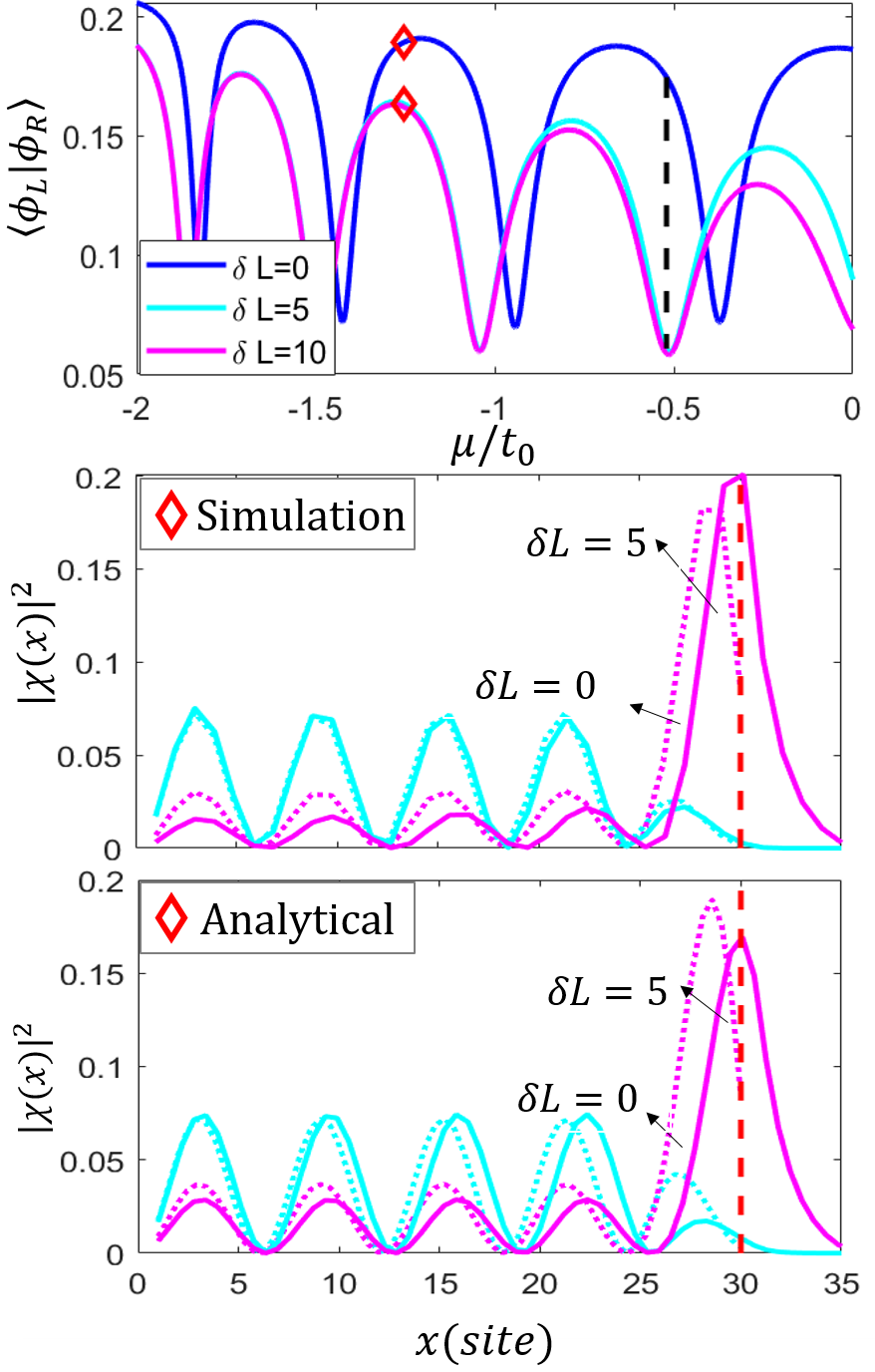}\llap{\parbox[b]{170mm}{\large\textbf{(a)}\\\rule{0ex}{130mm}}}\llap{\parbox[b]{170mm}{\large\textbf{(b)}\\\rule{0ex}{86mm}}}\llap{\parbox[b]{170mm}{\large\textbf{(c)}\\\rule{0ex}{43mm}}}
	\end{center}
	\caption{(Color online){(a) Overlap of the left and right MBSs $\braket{\phi_L|\phi_R }$ of a partially separated ABS as a function of chemical potential $\mu$ for a Kitaev chain of varying length $\delta_L$ coupled to a QD which with a finite proximitized region of length $\delta_x=5$. Increasing the length of the Kitaev chain $\delta_L$ decreases the overlap between the left and right modes. The overlap value $0.1748$ is significantly reduced to $0.06$ for $\mu$ marked by the black dashed line. (b)-(c) Majorana wave functions for the lowest energy modes of a QD-Kitaev chain associated with red diamonds in (a) for numerical simulation (b) and analytical results based on Eq.~\ref{eq:dotTopVars} (c). Significant portions of the mode distribution leak into the Kitaev chain region $\delta_L$, reducing the overlap between the left and right modes, increasing the robustness of the ps-ABS. Here $t=5t_0, x_0 =30, \Delta=t_0, V=-2.5t_0$ were used.}
	}\label{fig:boundStatesOverlap}	
\end{figure}
\begin{figure}
	\begin{center}
		\includegraphics[width=0.45\textwidth]{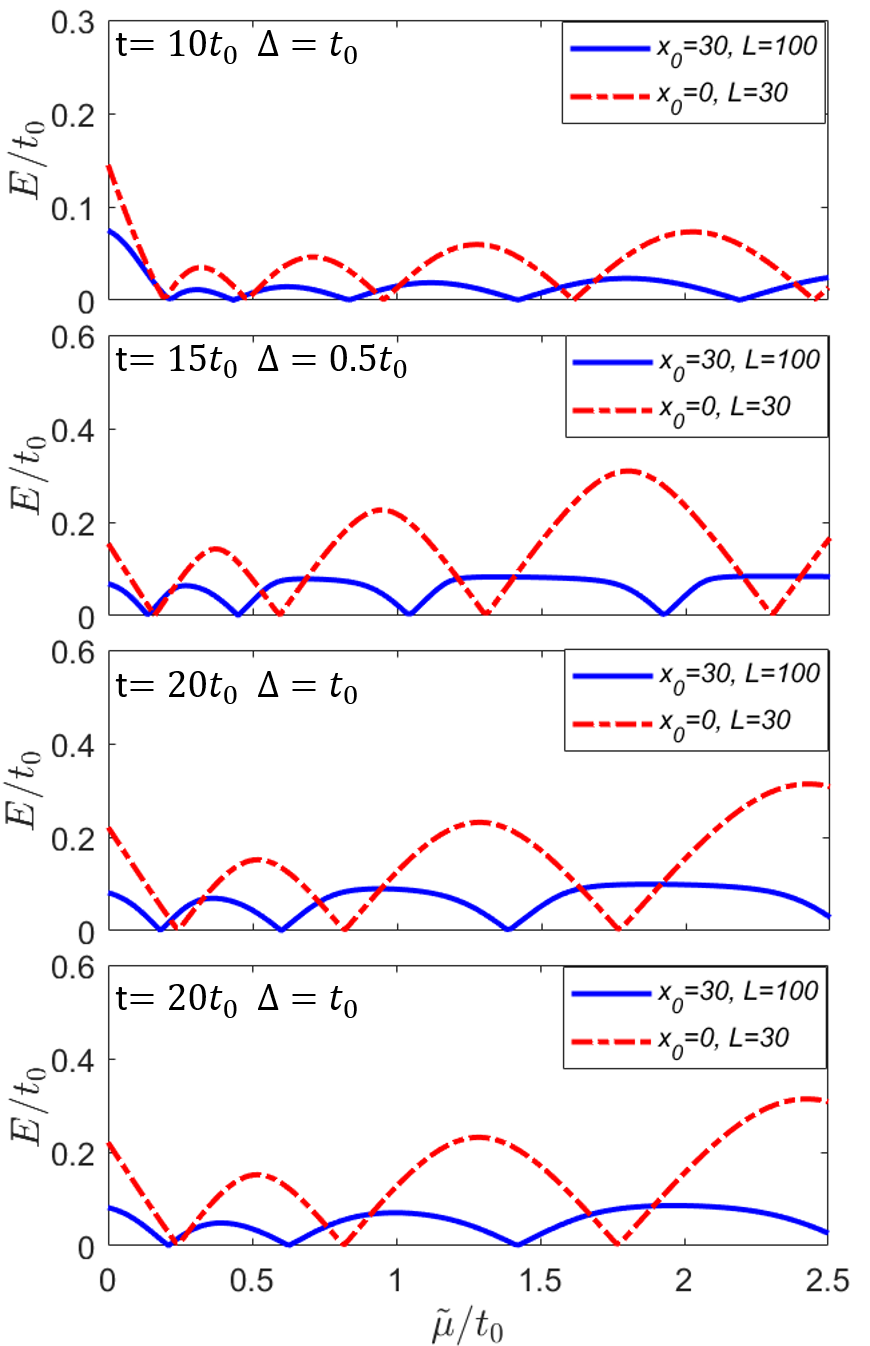}
		\llap{\parbox[b]{163mm}{\large\textbf{(a)}\\\rule{0ex}{120mm}}}
		\llap{\parbox[b]{164mm}{\large\textbf{(b)}\\\rule{0ex}{90mm}}}
		\llap{\parbox[b]{166mm}{\large\textbf{(c)}\\\rule{0ex}{61mm}}}
		\llap{\parbox[b]{167mm}{\large\textbf{(d)}\\\rule{0ex}{32mm}}}
	\end{center}
	\caption{(Color online) Lowest energy eigenvalues plotted with effective chemical potential for a Kitaev chain with QD in the topologically trivial phase (solid blue) and bare Kitaev chain of length equal to the QD in the topological phase (dashed red) with different parameters $t,\Delta,$ and the length of the proximitized region $\delta_x$. The parameters $x_0=30, L=100$ indicate a dot-chain system with a total length of $100$ sites attached to a QD with $30$ sites with the effective chemical potential in the QD $\tilde{\mu}=\mu-V$. The parameters $x_0=0, L=30$ indicate a bare Kitaev chain of length $30$ sites (and no QD), with the effective chemical potential $\tilde{\mu}=\mu$ ($V=0$ in the bare chain). We fix $\mu=-0.05t_0$ for the dot-chain system and vary the dot potential $V$ in a range such as $\tilde{\mu} > 0$ within the QD, which varies within the same range as that for the bare Kitaev chain. The lowest energy in the topologically trivial ($\mu <0$) dot-chain system (i.e., the ps-ABS) shows a significantly suppressed energy splitting as compared to that of the topological ($\tilde{\mu}=\mu>0$) Kitaev chain of length $x_0$. We have length of the proximitized region $\delta_x=10$ in (a)-(c), and $\delta_x=30$ in (d). Parameters $t,\Delta$ are as given, and $V=-2.5 t_0$.}\label{fig:ps-ABS-mzm}
\end{figure}
The lowest energy BdG wave functions based on Eq.~\ref{eq:absWavefunction}, shown in Fig.~\ref{fig:wfDot}(a), illustrate the critical importance of the proximitized region within the QD. When the effective chemical potential within the proximitized region $\tilde{\mu}\gtrapprox 0$, the solution given in Eq.~\ref{eq:waveFunction} is used, implying the formation of a pair of MBSs at the boundaries of the proximitized region. One of this pair of component Majorana bound states can ``leak'' into the normal part of the QD, while the other bound state remains localized within the QD, effectively separating the MBSs. When the MBSs are separated on the order of the characteristic energy decay length $\zeta\propto q_F^{-1}$ (as defined in Eq.~\ref{eq:energyDecay}) they form a ps-ABS \cite{moore2018two-terminal,moore2018quantized}  as shown in Fig.~\ref{fig:wfDot}(b)--(d) (where only the first 60 sites of the QD-chain is shown). We now define the overlap between the pair of component MBSs in terms of the spatial integral of the product of the absolute values of the wave functions,
$\braket{\phi_L|\phi_R}=\int dx |\phi_L||\phi_R|$.
 Plotting this overlap $\braket{\phi_L|\phi_R}$ as a function of the length of the proximitized region $\delta_x$ as in Fig.~\ref{fig:overlap} shows that if $\delta_x=0$, as shown in Fig.~\ref{fig:overlap}(a), there is a strong overlap (in red) throughout the topologically trivial region ($\mu<0$), signaling the presence of an ABS comprised of a pair of strongly overlapping MBS. On the other hand a proximitized region of finite length $\delta_x\gtrapprox 5$ within the QD allows for the formation of a robust low overlap (in blue) region, even in the topologically trivial regime, signaling the presence of a ps-ABS. As the length of the proximitized region $\delta_x$ increases, the overlap between the left and right MBSs comprising a ps-ABS decreases exponentially (Fig.~\ref{fig:overlap}(c)) even when the bulk of the Kitaev chain is in the topologically trivial regime.

For a partially proximitized QD of length $x_0=30$ attached to a Kitaev chain of length $\delta_L$, in which the effective potential within the proximitized region of the dot is $\tilde{\mu}\gtrapprox 0$ and the Kitaev chain is topologically trivial, the overlap between the left and right MBSs $\braket{\phi_L|\phi_R}$ decreases with increasing length of the Kitaev chain $\delta_L$, due to a portion of one of the component MBS leaking into the Kitaev chain.
Fig.~\ref{fig:boundStatesOverlap} shows results for a partially proximitized ($0 < \delta_x<x_0$) QD of length $x_0=30$ attached to a Kitaev chain of length $\delta_L$, in which the effective potential within the proximitized region of the dot is $\tilde{\mu}\gtrapprox 0$ and the Kitaev chain is topologically trivial ($\mu<0$). As shown in Fig.~\ref{fig:boundStatesOverlap}(a), the overlap $\braket{\phi_L|\phi_R}$ decreases with increasing length of the Kitaev chain $\delta_L$, owing to the fact that one of the component MBS of the ps-ABS can relax into the topologically trivial Kitaev chain. In Fig.~\ref{fig:boundStatesOverlap}(a) three different $\delta L$ for $t=5t_0$ and $x_0=30$ are analyzed, showing a reduction in overlap between the left and right MBSs comprising a ps-ABS with increasing length of $\delta L$. The oscillation of $\braket{\phi_L|\phi_R}$ with $\mu$ can be attributed to the oscillation of the wave functions ($\sim \sin(k_F x)$) when the boundary conditions are matched at $x=\left(x_0-\delta_x\right)$ and $x=x_0$. The reduction in overlap between the left and right MBSs is less prevalent between $\delta L =5$ and $\delta L=10$ than between $\delta L=0$ and $\delta L=5$, signaling that only the part of the Kitaev chain adjacent to the QD-Kitaev chain interface controls the relaxation of the MBS, and the progressive decrease of the wave function overlap in the ps-ABS, as expected. Analytical results for the square of the absolute values of the MBSs associated with the red diamonds in Fig.~ \ref{fig:boundStatesOverlap}(a) are shown and compared to numerical simulation in Fig.~ \ref{fig:boundStatesOverlap}(b)--(c), where a significant fraction of the probability density is shown to leak into the superconducting region of the Kitaev chain. This leads to a lower overlap between the left and right MBSs, decreasing the amplitudes of the splitting oscillations in ps-ABS compared to those for topological MZMs for an equivalent bare (without the quantum dot) Kitaev chain of length $x_0$, as shown in Fig.~\ref{fig:ps-ABS-mzm}.

\section{Summary and Conclusion}
In this paper, we have analytically solved the problem of a finite-length Kitaev chain coupled to a quantum dot, which, in addition to being a valuable extension of the classic Kitaev chain problem, is an effective representation of a system investigated in recent Majorana experiments: a spin-orbit coupled quandum dot-semiconductor-superconductor hybrid nanowire in the presence of a Zeeman field. Here, the quantum dot is defined by a portion of the SM wire not covered by the epitaxial SC, which can be under an electric potential controlled using external gates. Previously, we modeled such a QD in terms of a vanishing SC pair potential $\Delta$ and a step-like barrier potential $V$, \cite{moore2018two-terminal,moore2018quantized} which led us to the important result that robust, near-zero-energy, subgap Andreev states are the generic low energy excitations localized in the QD region in the topologically trivial phase of the SM wire. The assumption of a step-like barrier potential in the quantum dot region produces an effective potential profile that is manifestly different from the smooth confinement potential at the end of the SM-SC system as considered in Ref.~[\onlinecite{kells2012near}]. Specifically, while the pair of component MBSs that constitute a robust near-zero-energy ABS in the presence of smooth confinement potential \cite{kells2012near} originate from two \textit{different} spin channels of a confinement-induced sub-band, in the case of a potential barrier (or a potential well) in the QD the component MBSs originate from the \textit{same} spin channel. Consequently, while the topological properties of the QD-SM-SC hybrid structure with local step-like dot potential \cite{moore2018two-terminal,moore2018quantized} can be understood using an effective representation in terms of a Kitaev chain (which has a single spin channel) coupled to a QD, as discussed in the present work, the SM-SC heterostructure with smooth confinement potential \cite{kells2012near}  cannot be analyzed within such a representation.

Our key analytical result for the Kitaev chain coupled to a QD is demonstrating the existence of a robust near-zero-energy ABS (localized in the QD region) in the topologically trivial phase ($\mu<0$) of the Kitaev chain. By contrast, topological near-zero-energy MZMs separated by the chain length $L$ are the lowest energy excitations in the topological superconducting phase ($\mu>0$) of the Kitaev chain. Our analysis reveals the crucial importance of a slice of the QD being proximitized, which may correspond in the experiments to the potential barrier slightly penetrating into the region covered by the SC. We show that only in the presence of such a slice of proximitized region in the QD, the eigenvalue equation and the boundary conditions admit a robust, near-zero-energy, subgap ABS in the topologically trivial phase of the Kitaev chain. Furthermore, the component pair of MBSs of this topologically trivial ABS are spatially separated by the width of the proximitized part of the QD, leading to the so-called partially separated ABSs (ps-ABS) and the resultant robustness to local perturbations of the zero bias conductance peaks in tunneling measurements,\cite{moore2018quantized} as seen in the experiments.\cite{zhang2018quantized}

The analytical calculations also reveal that the ps-ABSs appear whenever the effective chemical potential in the proximitized part of the QD $\tilde{\mu}=\mu - V \gtrapprox 0$, allowing the partial decoupling of the component MBSs and nucleating a ps-ABS in the QD, even though the bulk of the Kitaev chain may be in the trivial phase $\mu < 0$. In the present case, this requires a potential well $V<0$ in the QD near the Kitaev chain TQPT at $\mu=0$. Near the Kitaev chain TQPT at $\mu=4t$, the conditions $\tilde{\mu}=\mu - V \lessapprox 4t$ and $\mu > 4t$ in the bulk of the chain require the presence of a potential barrier ($V > 0$) at the QD. In the analogous spin-full problem of the QD-SM-SC heterostructure the nucleation of a ps-ABS in the proximitized part of the QD can take place in the presence of either a potential well ($V<0$), or  a potential barrier ($V>0$), but the separation of the component MBSs (hence, the robustness of the ps-ABS) is typically stronger for $V>0$.\cite{moore2018two-terminal,moore2018quantized}. Finally, we also find the important result that the energy splittings in the ps-ABS are significantly suppressed than the energy splittings expected in a bare topological segment of equivalent length (typically, the size of the QD), because the component MBS of a ps-ABS localized near the QD-SC interface can relax into the adjacent Kitaev chain which is in the topologically trivial phase.

\section{Acknowledgments}
C.M., C.Z., and S.T. acknowledge support from ARO Grant
No. W911NF-16-1-0182. T.D.S. was supported by NSF Grant
No. DMR-1414683.

%
\end{document}